# Modulation Instability of Ultrashort Pulses in Quadratic Nonlinear Media beyond the Slowly Varying Envelope Approximation

Amarendra K. Sarma* and Parvendra Kumar
Department of Physics, Indian Institute of Technology Guwahati, Guwahati-781039, Assam, India.
*Electronic address: aksarma@iitg.ernet.in

**Abstract:** We report a modulational instability (MI) analysis of a mathematical model appropriate for ultrashort pulses in cascaded quadratic-cubic nonlinear media beyond the so-called slowly varying envelope approximation. The study shows the possibility of controlling the generation of MI and formation of solitons in a cascaded quadratic-cubic media in the few cycle regimes.

## 1. Introduction

Recently nonlinear optics research beyond the so called slowly varying envelope approximation (SVEA) have received tremendous boost due to various reasons, primarily for richness in physics and possible applications in many diverse areas such as, ultrafast spectroscopy, metrology, medical diagnostics and imaging, optical communications, manipulation of chemical reactions and bond formation, material processing etc. [1-2]. Particularly, the availability of sources of light in the near-single optical cycle has opened new possibilities for physicists and scientist to explore and doubt many of the fundamental concepts and assumptions [3-4].The validity of SVEA is already questioned by many authors in this new domain of optical science [5-10].Many authors have attempted to modify the SVEA so that it might be extended to the few cycle regimes. The first widely accepted model in this regard has been developed by Brabec and Krausz [5]. Some other authors have offered non-SVEA models also [11-12]. However, the model equation proposed by Brabec and Krausz have been used most extensively and successfully in various contexts [13-16]. Recently, following the model proposed by Brabec and Krausz, Moses and Wise have derived a coupled propagation equations for ultrashort pulses in a degenerate three-wave mixing process in quadratic ($\chi^{(2)}$) media [17]. In passing, it is worthwhile to mention that owing to the efficient manipulation of spectral and temporal properties of femtosecond pulses through cascaded processes in quadratic materials, both theoretical and experimental research is getting tremendous boost in recent years [18-24]. We should also note that Moses-Wise model is restricted to the case of strongly mismatched interaction where the conversion efficiency to second or higher harmonics is negligible. In fact, a more generalized nonlinear envelope equation is derived by Conforti et al. to describe the propagation of broadband optical pulses in second order nonlinear materials [22-23]. However, Moses and Wise went on to present, using cascaded quadratic nonlinearity, theoretical and experimental evidence of a new quadratic effect, namely the controllable self-steepening (SS) effect. The controllability of the SS effect is very useful in nonlinear propagation of ultrashort pulses as it may be used to cancel the propagation effects of group velocity mismatch. After publication of the Moses-Wise model it is pointed out that the cascaded nonlinearity induces a Raman-like term into the model owing to the nonlocal nature of the cascaded nonlinearity [24]. It may be noted that, traditionally, the intensity dependent refraction (IDR) effects in quadratic media are not expected in quadratically nonlinear media owing to the phase mismatch of the fundamental harmonic with the higher ones within the SVEA [25]. In this work, we have studied the modulation instability (MI) of the single-field equation for the fundamental field (FF) derived by Moses and Wise [17]. Our study is mainly motivated by the fact that IDR effect is closely related to MI, particularly to the existence of optical solitons in a nonlinear media. It is well known that modulation instability is a fundamental and ubiquitous process that appears in most nonlinear systems in nature [26-31]. It occurs as a result of interplay between the nonlinearity and dispersion in time domain or diffraction in spatial domain. It may be noted that the modulational instability in a quadratic media was first studied by Trillo and Ferro taking into account the parametric interaction between fundamental and second harmonic field in a transparent dispersive medium [32-32*]. However, cascading in quadratic media and solitary wave propagation was first discussed in the work of Menyuk, Schiek and Torner [33].Later, MI in quadratic media is explored by many authors, in various context such as competition between spatial and temporal break up[34], quantum-noise-initiated break up[35] etc. In this work we are specifically interested to see the role of group velocity mismatch (GVM) between the fundamental (FF) and second-harmonic (SH) field on MI as well as the role of self-steepening (SS) parameter. Though our study indicates the possibility of getting MI even in the so-called normal dispersion regime, in this work we mainly confine our attention to the anomalous dispersion regime only.

## 2. Theoretical model and MI analysis

The Moses-Wise model for ultrashort pulse propagation in a cascaded-quadratic media could be written as follows [17]:

$$i\frac{\partial A}{\partial z} - \frac{\beta_2}{2}\frac{\partial^2 A}{\partial T^2} - i\frac{\beta_3}{6}\frac{\partial^3 A}{\partial T^3} + \left(\gamma + \frac{\Gamma^2}{\Delta k}\right)|A|^2 A + i\left(\frac{\Gamma^2}{\Delta k \omega_0} + \frac{\gamma}{\omega_0}\right)A^2\frac{\partial A^*}{\partial T}$$
$$+ i\left(-\frac{2\delta\Gamma^2}{\Delta k^2} + \frac{3\Gamma^2}{\Delta k \omega_0} + \frac{2\gamma}{\omega_0}\right)|A|^2\frac{\partial A}{\partial T} = 0 \tag{1}$$

Here $A$ is the complex envelope of the fundamental field travelling along $z$, $\beta_2$ is the group velocity dispersion(GVD) parameter, $\beta_3$ is the third order dispersion (TOD) parameter, $\omega_0$ is the carrier wave frequency, $\gamma$ is the cubic nonlinear co-efficient, $\Delta k = 2k_1 - k_2$ is the wave vector mismatch between the FF and SH fields and $\Gamma^2 = 32\pi^2\omega_0^4\chi^{(2)}_{2\omega_0-\omega_0}\chi^{(2)}_{\omega_0+\omega_0}/k_1k_2c^4$. $k_1$ and $k_2$ are respectively the wave vectors associated with FF and SH fields. $\delta$ measures group velocity-mismatch between FF and SH fields. It may be worthwhile to mention a few words about the fifth term of Eq. (1). Here the first term inside the bracket refers to SS arising from $\chi^{(2)}$ while the second one refers to SS arising from $\chi^{(3)}$ [17]. GVM-induced SS term is included in the last term through the parameter $\delta$. It may be noted that the fourth and the fifth term of Eq. (1) vanishes if $\Gamma^2/\Delta k = -\gamma$, while the last term could vanish for some particular values of $\Delta k$ and $\gamma + \Gamma^2/\Delta k \neq 0$ as discussed in Ref.[17]. For a systematic derivation of the Moses-Wise model readers are referred to Ref. [36] along with Ref. [17].Now we rewrite Eq. (1) in the so called soliton units [26] as follows:

$$i\frac{\partial u}{\partial \xi} - \frac{\alpha}{2}\frac{\partial^2 u}{\partial \tau^2} - i\delta_3\frac{\partial^3 u}{\partial \tau^3} + \beta|u|^2 u + i\beta s u^2\frac{\partial u^*}{\partial \tau} + i\rho|u|^2\frac{\partial u}{\partial \tau} = 0 \tag{2}$$

where $u$ is the normalized amplitude, $\alpha = \text{sgn}(\beta_2)$ and

$$\xi = z/L_D, \tau = T/T_0, L_D = T_0^2/|\beta_2|, A = \sqrt{P_0}U, u = NU, N = \sqrt{\gamma P_0 L_D}, \delta_3 = \beta_3/6|\beta_2|T_0,$$
$$\beta = 1 + (\Gamma^2/\Delta k \gamma), s = 1/\omega_0 T_0, \rho = 2s + 3s(\Gamma^2/\Delta k \gamma) - 2(\Gamma^2\delta/\Delta k^2 T_0 \gamma) \tag{3}$$

Here $\xi$ and $\tau$ are the normalized propagation distance and time respectively, $P_0$ is the peak power of the incident pulse, $L_D$ is the dispersion length and $N$ is the so called soliton order. We would like to emphasise that in this work, $s$ is termed as the normalized self-steepening (SS) parameter in analogy with the one found in conventional nonlinear fibre optics [26]. On the basis of Eq. (2) we would now investigate the MI of few cycle pulses. Eq. (2) has a steady state solution given by $u = u_0 \exp\left[i\beta u_0^2 \xi\right]$, where $u_0$ is the constant amplitude of the incident plane wave. We now introduce perturbation $a(\xi,\tau)$ together with the steady state solution to Eq. (2) and linearize in $a(\xi,\tau)$ to obtain:

$$i\frac{\partial a}{\partial \xi} + \beta u_0^2 (a + a^*) - \frac{\alpha}{2}\frac{\partial^2 a}{\partial \tau^2} - i\delta_3 \frac{\partial^3 a}{\partial \tau^3} + i\beta s u_0^2 \frac{\partial a^*}{\partial \tau} + i\rho u_0^2 \frac{\partial a}{\partial \tau} = 0 \qquad (4)$$

Separating the perturbation to real and imaginary parts, according to $a = a_1 + i a_2$, and assuming $a_1, a_2 \propto \exp\left[i(K\xi - \Omega\tau)\right]$, where $K$ and $\Omega$ are the wave number and the frequency of perturbation respectively, from Eq. (4) we obtain the following dispersion relation

$$K = \Omega^3 \delta_3 + \rho u_0^2 \Omega \pm \left[\frac{\Omega^4}{4} + \left(1 + \frac{\Gamma^4}{\Delta k^2 \gamma^2}\right) s^2 u_0^4 \Omega^2 + \frac{\Gamma^2}{\Delta k \gamma}\left(2 s^2 u_0^4 \Omega^2 + \alpha \Omega^2 u_0^2\right) + \alpha \Omega^2 u_0^2\right]^{\frac{1}{2}} \qquad (5)$$

From Eq. (5), we observe that the modulation instability exists only if the quantity inside the bracket is <0 and TOD plays no role in MI. The fact that third-order dispersion contributes nothing to modulational instability is a well-known result [37-38]. Another interesting fact is that, the sixth term of Eq. (1) plays no role in MI.

To start with the MI analysis, we begin with the case of anomalous dispersion regime, for which $\alpha = -1$. The expression for the so-called gain spectrum $g(\Omega)$ could be put in the following form:

$$g(\Omega) = 2\,\text{Im}(K) = 2\left[\Omega_B^2 - \Omega_A^2\right]^{\frac{1}{2}}, \qquad (6)$$

where,

$$\Omega_B^2 = \Omega^2 u_0^2 - \eta\left(2 s^2 u_0^4 - u_0^2\right)\Omega^2; \quad \Omega_A^2 = (1+\eta^2) s^2 u_0^4 \Omega^2 + (\Omega^4/4) \quad \text{with } \eta = \Gamma^2/\gamma\Delta k \qquad (7)$$

We note that for the occurrence of MI, one must have $\Omega_B^2 > \Omega_A^2$. The maximum of the gain occurs at two frequencies given by:

$$\Omega_{\max} = \pm\sqrt{2u_0^2 - 2\eta\left(2 s u_0^4 - u_0^2\right) - 2(1+\eta^2) s^2 u_0^4} \qquad (8)$$

Now, we would try to see the role of various controllable parameters such as the self-steepening (SS) $s$ and wave-vector mismatch $\Delta k$ on MI. Fig.1 depicts the modulation instability gain spectrum $g(\Omega)$ vs. $\Omega$ for different values of $\eta$ for $s = .01$ and $u_0 = 1$.

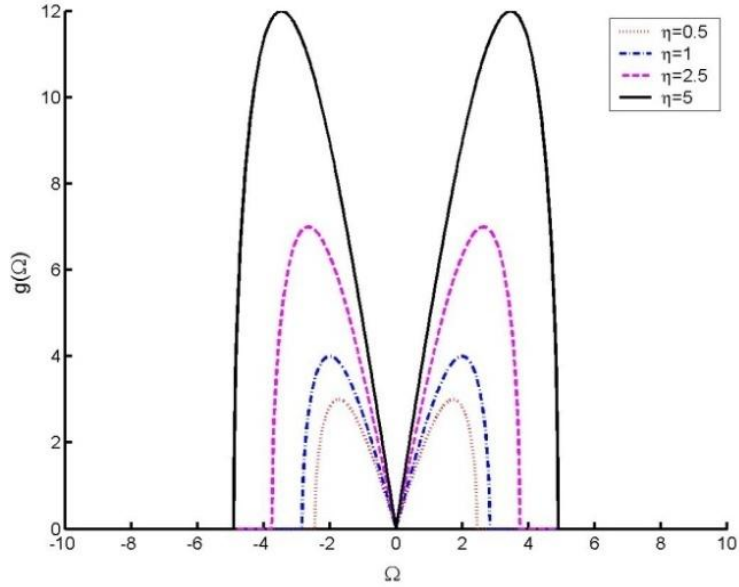

Fig. 1 (Color online) Modulational instability gain as a function of normalized frequency for four different values of $\eta$ with $u_0=1$ and $s=0.1$.

It can be clearly seen that the gain spectrum is symmetric with respect to $\Omega=0$. We observe from Fig. 1 that, for the given input power and a fixed self-steepening parameter, the modulation instability gain increases with increase in $\eta$. Physically speaking, MI gain increases with a decrease in the wave-vector mismatch $\Delta k$, as the parameter $\eta$ is directly related to it through Eq. (7). If a probe wave at a frequency $\omega_0+\Omega$ were to propagate with the CW beam at $\omega_0$, it would experience a net power gain given by Eq.(6) as long as $\text{Im}(K)>0$. Eventually, due to MI gain, the CW beam would break up spontaneously into a periodic pulse train known as solitons. These soliton-like pulses exist whenever the conditions $\Omega_B^2>\Omega_A^2$ and $\Delta k \neq 0$ are satisfied. The appearance of the sidebands located around $\Omega=0$ is the clear evidence of modulation instability.

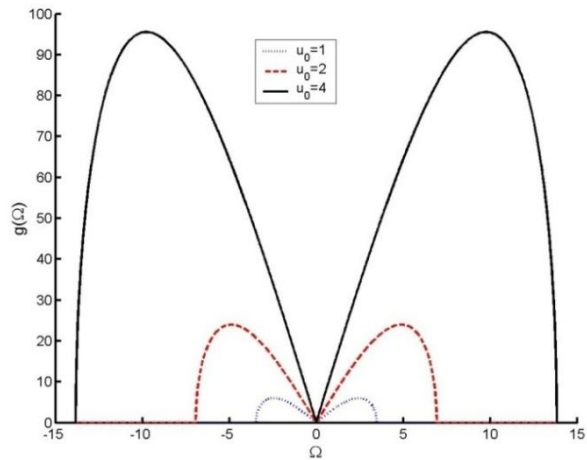

Fig. 2 (Color online) Modulation instability gain as a function of normalized frequency for three different values of $u_0$ with $\eta=2$ and $s=0.1$.

Fig.2 shows, quite expectedly, that with increase with the amplitude, the MI gain also increases. On the other hand we find that MI gain decreases with increase in the SS parameter, $s$, as could be observed from Fig.3. As the SS parameter is inversely related to both the carrier-wave frequency, $\omega_0$, and pulse width, $T_0$, we may conclude that through judicious choice of parameters one may control the modulational instability gain, in a cascaded quadratic media like the one considered in this work. It may be possible to have MI in the normal dispersion regime, for which $\alpha = +1$, if $\Delta k < 0$ and other parameters are chosen judiciously. This would necessitate tuning the phase mismatch to make the cascaded nonlinearity stronger than the Kerr one. So, finally to get an idea about the occurrence of the modulation instability in the normal dispersion regime, for which $\alpha = +1$, in Fig. 4 we depict MI gain as a function of the normalized frequency for three different values of $\eta$ with $s = 0.01$ and $u_0 = 1$.

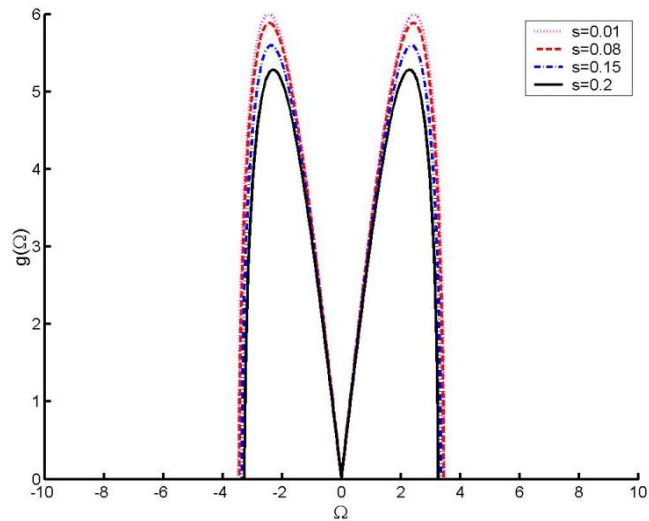

Fig. 3 (Color online) Modulation instability gain as a function of normalized frequency for four different values of $s$ with $\eta = 2$ and $u_0 = 1$.

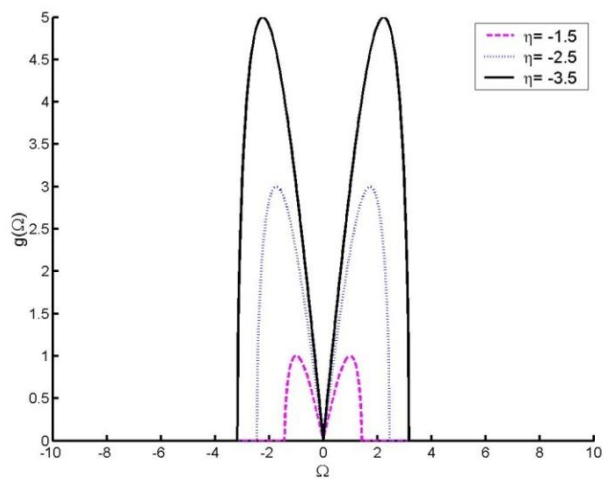

Fig. 4 (Color online) Modulational instability gain as a function of normalized frequency for three different values of $\eta$ with $u_0 = 1$ and $s = 0.01$ (Normal dispersion regime)

We note that the parameter $\eta$ and thereby the $\beta$-parameter is now negative. As stated earlier, this would require one to tune the phase mismatch so that the cascaded nonlinearity becomes stronger than the Kerr one. MI gain increases with increase in $|\beta|$. Similar analysis for other parameters may be carried out as was done for the case of anomalous dispersion regime.

### 3. Conclusion

To conclude, we have studied the modulation instability of the Moses-Wise model for ultrashort pulse propagation in a cascaded-quadratic media. A nonlinear dispersion relation is worked out using standard methods. We find that subject to the fulfilment of the MI criteria and judicious choice of the parameters, MI could be generated in a cascaded quadratic-cubic media in both normal and anomalous dispersion regimes.

**Acknowledgement**

A.K.S. sincerely thanks the Department of Science and Technology, Government of India for a research grant (grant no. SR/FTP/PS-17/2008).